\documentclass[twocolumn, amsmath, amssymb, aps, prl]{revtex4-1}

\usepackage{graphicx}
\usepackage{pdfpages}
\usepackage{eso-pic}
\usepackage{everyshi}
\usepackage{multido}

\begin{document}

\title{CO Tip Functionalization Inverts Atomic Force Microscopy Contrast via Short-Range Electrostatic Forces}

\author{Maximilian Schneiderbauer}
\email{maximilian.schneiderbauer@ur.de}
\author{Matthias Emmrich}
\author{Alfred J. Weymouth}
\author{Franz J. Giessibl}
\affiliation{Institute of Experimental and Applied Physics, University of Regensburg, 93040 Regensburg, Germany}
\date{February, 2014}

\begin{abstract}
We investigated insulating Cu$_2$N islands grown on Cu(100) by means of combined scanning tunneling microscopy and atomic force microscopy with two vastly different tips: a bare metal tip and a CO-terminated tip. We use scanning tunneling microscopy data as proposed by Choi \textit{et al.} \cite{Choi2008} to unambiguously identify atomic positions. Atomic force microscopy images taken with the two different tips show an inverted contrast over Cu$_2$N. The observed force contrast can be explained with an electrostatic model, where the two tips have dipole moments of opposite directions. This highlights the importance of short-range electrostatic forces in the formation of atomic contrast on polar surfaces in non-contact atomic force microscopy.
\end{abstract}

\maketitle

The combination of scanning tunneling microscopy (STM) with non-contact atomic force microscopy (NC-AFM) in a single probe enables a wide range of atomic scale studies on surfaces. Whereas contrast mechanisms in STM for different tip-sample systems are widely understood, interpretation of NC-AFM data remains challenging. In NC-AFM the sum over all tip-sample interactions is measured, and the source of atomic resolution is often hard to identify. On semiconductors \cite{Perez1997}, as well as on metals \cite{Dieska2003}, imaged with reactive tips (e.g. Si) atomic contrast is dominated by the formation of covalent bonds that often reach magnitudes of nanonewtons. For non-reactive CO functionalized tips, Pauli repulsion was attributed to the observed intermolecular resolution \cite{Gross2009, Moll2010}. Lantz \textit{et al.} \cite{Lantz2003} showed that the dangling bonds of Si(111) 7$\times$7 can induce a dipole moment in (non-reactive) oxidized Si tips resulting in a short-range electrostatic interaction, which contributes to atomic resolution. Electrostatic interaction and an induced tip dipole moment was also used to explain atomic contrast on ionic crystals \cite{Giessibl1992}. A similar model describes the interaction with charged adatoms on thin insulating layers \cite{Gross2009a, Bocquet2011}. Moreover, it was found that clean metallic tips carry an intrinsic dipole moment \cite{Teobaldi2011, Trevethan2012}, which is caused by the Smoluchowski-effect \cite{Smoluchowski1941}. All of these examples underline the importance of atomic scale electrostatic interactions in NC-AFM. 

\begin{figure}
\includegraphics[width=\columnwidth]{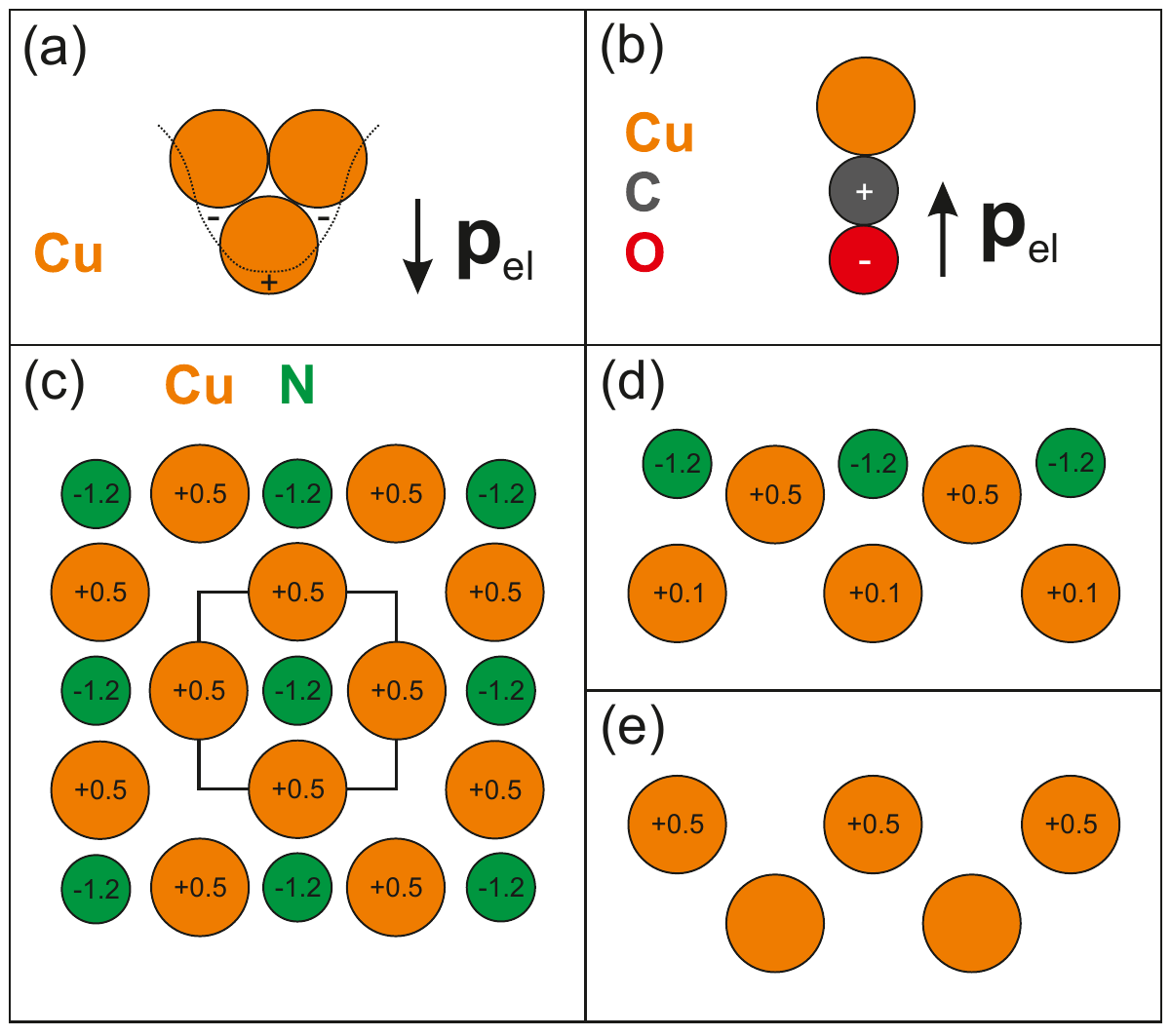}
\caption{(a) Charge redistribution in a metallic tip due to the Smoluchowski effect, leading to an electric dipole moment pointing towards the sample. (b) A CO molecule adsorbed on a copper tip, carrying an electric dipole moment pointing away from the sample. (c) Model of the Cu$_2$N network on Cu(100) with the numbers being the net charge of Cu and N atoms in units of elementary charge $e$ \cite{Hirjibehedin2007}. (d) Cross section of Cu$_2$N along a N-Cu-N row and (e) Cu-hollow site-Cu row showing the first two atomic layers of the reconstructed surface and the net charges of the atoms respectively.}
\label{Fig1}
\end{figure}

Electrostatic forces become even more meaningful as polar thin insulating layers (e.g. NaCl, MgO, Cu$_2$N) are used to decouple adsorbates in STM and AFM experiments \cite{Repp2004, Repp2005, Gross2009, Gross2009a, Sterrer2007, Hirjibehedin2006, Loth2010}. In this study we explore the influence of electrostatic forces in NC-AFM on Cu$_2$N islands on Cu(100). N and Cu atoms on Cu$_2$N form a periodic charge arrangement, as calculated by DFT \cite{Hirjibehedin2007} (Figs. \ref{Fig1}c-e). Compared to alkali halides, the Cu$_2$N's c(2$\times$2) unit cell structure has a lower symmetry, thus its atomic positions are easier to designate. STM experiments led to two criteria to locate N atoms within the islands \cite{Choi2008}: First, N adsorbs on the hollow sites of Cu(100) \cite{Leibsle1994, Yoshimoto2002, Soon2008} and should therefore appear fourfold symmetric. Second, island boundaries and sharp edges are determined by N atoms \cite{Hirjibehedin2006}. With this, N-, Cu- and hollow sites can be identified in the Cu$_2$N unit cell (Fig. \ref{Fig1}c).

In this Letter, we report on high-resolution simultaneously recorded current and force data of the Cu$_2$N surface. We compare interaction forces probed with a CO terminated tip to data acquired with a metal tip. The force contrast, though, is inverted. We propose an electrostatic model where the two tips have opposite dipole moments. The calculated force contrast within a Cu$_2$N unit cell provides good agreement to the data.

All experiments were carried out with a home-built low-temperature system at 6\,K using a qPlus sensor ($f_0 = 29098$\,Hz, $k=1800$\,Nm$^{-1}$) \cite{Giessibl2000} equipped with a W tip, operated with small amplitudes (50\,pm) in frequency modulation mode. Metallic (Cu) tips were prepared by strongly poking them into a clean Cu sample while applying 200\,V. Tips were functionalized with CO molecules following the standard procedure \cite{Bartels1997a}. Cu$_2$N islands were prepared by sputtering a clean Cu(100) crystal with N gas for 120\,s and heating it to 600\,K for 300\,s. Forces were calculated by applying the Sader-Jarvis-deconvolution method \cite{Sader2004} to our recorded 3D frequency shift maps. 

\begin{figure*}[t]
\includegraphics[width=1.6\columnwidth]{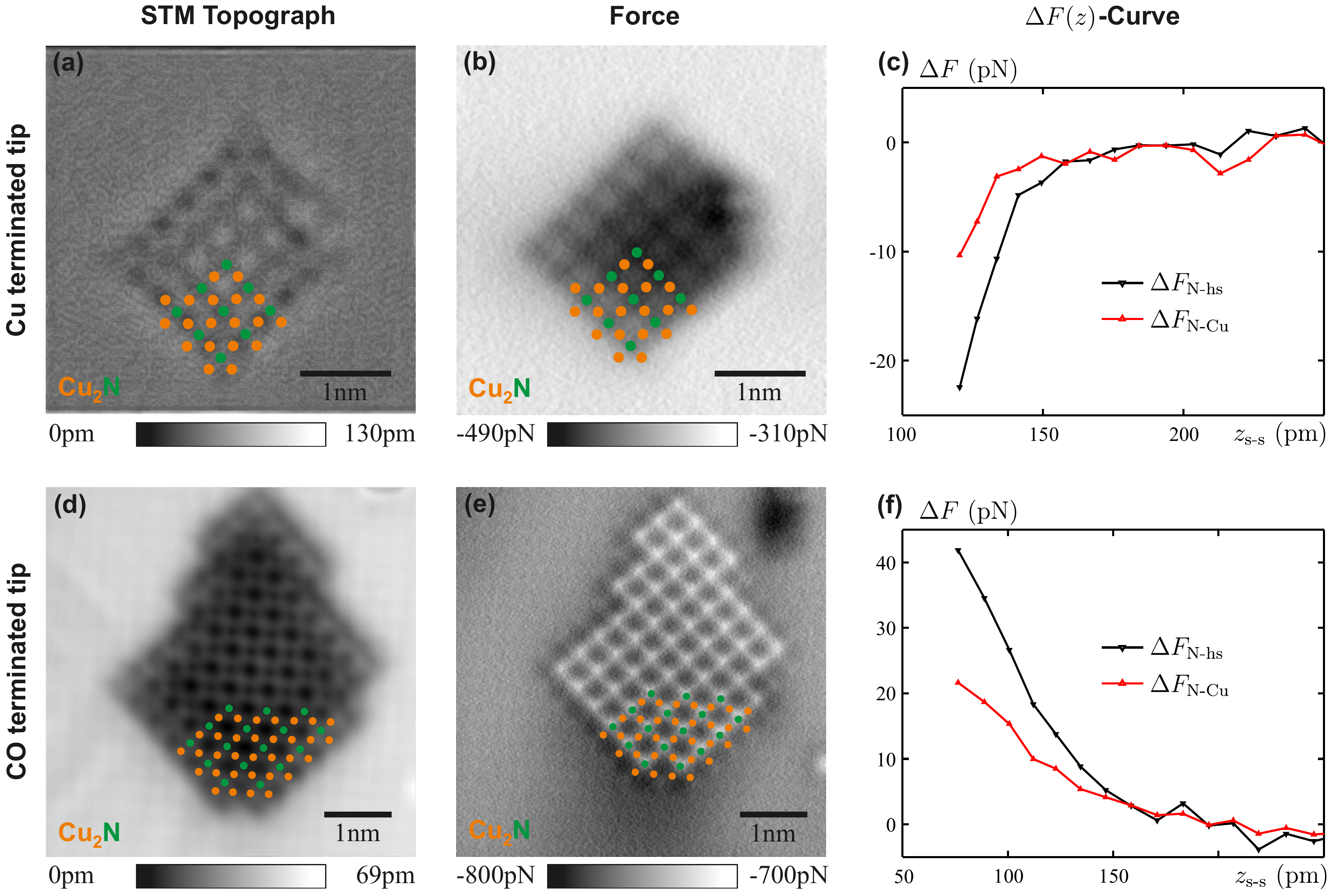}
\caption{Data acquired with a Cu terminated tip (a-c) and with a CO terminated tip (d-f). Left column: STM topographs with the structural model of Cu$_2$N overlaid: (a) is recorded at constant current at 500\,pA and 10\,mV applied to the tip and Laplace filtered to enhance contrast, (b) 100\,pA and 10\,mV, raw data. Center column: Force between tip and sample at closest approach. Right column: Force contrast versus distance curves at two high-symmetry locations.}
\label{Fig2}
\end{figure*}

Figure \ref{Fig2}a shows a STM topograph of an island using a Cu terminated tip recorded with a set point of 500\,pA at a tip voltage of 10\,mV. The image was Laplace filtered and inverted (Laplace filtering inverts brightness) to enhance contrast \cite{Hirjibehedin2006} and a structural model is partially overlaid.

When the Cu tip is terminated by a CO molecule, the STM contrast is different. The constant-current topograph in Fig. \ref{Fig2}d was recorded using a set point of 100\,pA at a tip voltage of 10\,mV. Three distinct features appear within the island: elongated protrusions, wide depressions and, most remarkable, very narrow depressions. It is known that the strong p-wave character of CO terminated tips influences the imaging contrast in STM \cite{Gross2011}. In Ref. \cite{Gross2011}, the authors investigated the organic molecule pentacene, whose orbitals locally exhibit $\sigma$, $\pi$ and $\delta$ character. They use Chen's derivative rule \cite{Chen1993} to explain that due to the CO's $\pi$ orbital the tunneling matrix element turns to zero over the pentacene's local $\sigma$ and $\delta$ orbitals, whereas not for the local $\pi$ orbital. We propose a similar tunneling contrast formation for copper nitride, motivated by the DFT calculations of Soon \textit{et al.} \cite{Soon2008}. For the eigenstate closest to our bias voltage, of 10\,mV, N and Cu atoms show a local $\sigma$ character, in which the N wave function has twice the spatial extension of Cu. Due to the 3\textit{d} state of Cu and the 2\textit{p} state of N, the space in between N and Cu shows a local $\pi$ character. Taking into account the N-Cu binding length of 183\,pm, which is comparable to atomic distances within pentacene, we argue as follows: N exhibits a local $\sigma$ character and therefore the matrix element concerning the CO's $\pi$ orbital is zero, resulting in a very confined depression over N. Between N and Cu the CO tip probes a local $\pi$ orbital enabling a tunneling current. Cu also has local $\sigma$ character, whereby its wave function only extends half compared to N. Furthermore, Cu is located 21\,pm lower than N \cite{Yoshimoto2002, Soon2008}. From this we conclude that our microscope setup is not able to laterally resolve a depression over Cu as well, resulting in elongated protrusions centered over Cu. For hollow sites, the lateral orbital overlap between tip and sample states is insufficient to yield a non-zero tunneling matrix element, leading to a wide depression. This lattice assignment fulfills the two required assignment criteria (Figs. \ref{Fig2}d-e).

Figure \ref{Fig2}b shows the total force for a Cu terminated tip at closest approach, corresponding to a STM set point of 28\,nA at a tip voltage of 10\,mV over bare Cu(100). Because the measured tunneling current $I$ is proportional to the conductance $G$, the STM set point can be given in units of the quantum conductance $G_0=2 e^2/h=(12,906\,\Omega)^{-1}$ ($e$ is the elementary charge and $h$ is the Planck constant). As $I$ depends exponentially on the tip-sample distance $z$ with decay constant $\kappa$, $G$ can be used to calculate the distance between the front most metallic tip and sample atom, with $G(z=0)=G_0$ \cite{Ternes2008}. This calculation yields a lower boundary for the tip-sample distance, as typically more than two atoms contribute to the measured tunneling current \cite{SM}. Here, tip-sample distance is defined as the distance between the outermost atomic shells of the involved atoms, in the following called $z_{\text{s-s}}$. Using this definition of distance, Fig. \ref{Fig2}b was measured at $z_{\text{s-s}}=125$\,pm over bare Cu(100).

According to the lattice assignment for metal tips, N sites appear most attractive, followed by Cu sites and hollow sites. The measured overall force is attractive, because of the long-range van-der-Waals interaction, which normally does not depend on lateral position and thus does not allow for atomic resolution. Subtracting the forces between distinct unit cell positions cancels out constant long-range interactions and the resulting force contrast contains site dependent short-range components only. We define $\Delta F_{\text{N-Cu}}=F_{\text{N site}}-F_{\text{Cu site}}$ and $\Delta F_{\text{N-hs}}=F_{\text{N site}}-F_{\text{hs}}$ as the force difference/contrast between N and Cu sites and between N sites and hollow sites (hs). Table \ref{table} shows the averaged force contrast for five different islands probed with different Cu terminated tips at similar tip-sample-distances. The short-range character of the atomic contrast causing interaction is reflected in the $\Delta F(z)$-curves displayed in Fig. \ref{Fig2}c. The atomic contrast is maintained over a vertical range of about 100\,pm.

\begin{table}
\renewcommand{\arraystretch}{1.5}
\begin{tabular}{l|c|c|c|c}
       & \multicolumn{2}{|c|}{$\Delta F_{\text{N-Cu}}$} &  \multicolumn{2}{|c}{$\Delta F_{\text{N-hs}}$} \\ 
			 & experiment      & simulation                   & experiment       & simulation \\ \hline
Cu tip & (-10$\pm$2)\,pN & -12\,pN                      & (-17$\pm$3)\,pN  & -17\,pN    \\ \hline
CO tip & (22$\pm$1)\,pN  & 26\,pN                       & (39$\pm$2)\,pN   & 34\,pN
\end{tabular}
\caption{Comparison of averaged experimental and simulated force contrast for Cu and CO terminated tips.}
\label{table}
\end{table}

Figure \ref{Fig2}e shows the total force at closest approach for a CO terminated tip, which was at a STM set point of 3\,nA at 10\,mV over clean Cu(100). To determine the tip-sample distance the above described model has to be adapted by using the point conductance of a CO molecule on the Cu surface. Experimentally we find $G_0=(404,497\,\Omega)^{-1}$ \cite{SM} and obtain $z_{\text{s-s}}=80$\,pm for the closest approach. Figure \ref{Fig2}e reveals an inverted force contrast compared to Fig. \ref{Fig2}b, with N atoms interacting most repulsive and hollow site most attractive. The averaged force contrast for three CO functionalized tips over three different islands at similar tip-sample-distances is depicted in Table \ref{table}. Figure \ref{Fig2}f displays the above defined $\Delta F(z)$-curves, showing repulsive atomic contrast for approximately 100\,pm.

The AFM images in the center column of Fig. \ref{Fig2} show a contrast inversion for Cu vs. CO tips. We attribute this contrast inversion to opposite dipole moments of Cu vs. CO terminated tips for two reason. First, Cu$_2$N is a periodic arrangement of charged atoms. The electrostatic potential of such a periodic charge distribution with lattice constant $a$ decays exponentially \cite{Lennard-Jones1928}, with a decay length given by $\lambda=a/2\pi$ (Cu$_2$N: $a=372$\,pm \cite{Choi2008} and thus $\lambda=59$\,pm). This small decay length explains the short-range character of this electrostatic interaction (Figs. \ref{Fig2}c,f). Second, in the following we will present an electrostatic model using a point-charge representation of sample and tip. Using the force contrast and calculated tip-sample distances from the experiment, we fitted the tip's dipole moment to model the experimental force dependence.

\begin{figure*}[t]
\includegraphics[width=1.6\columnwidth]{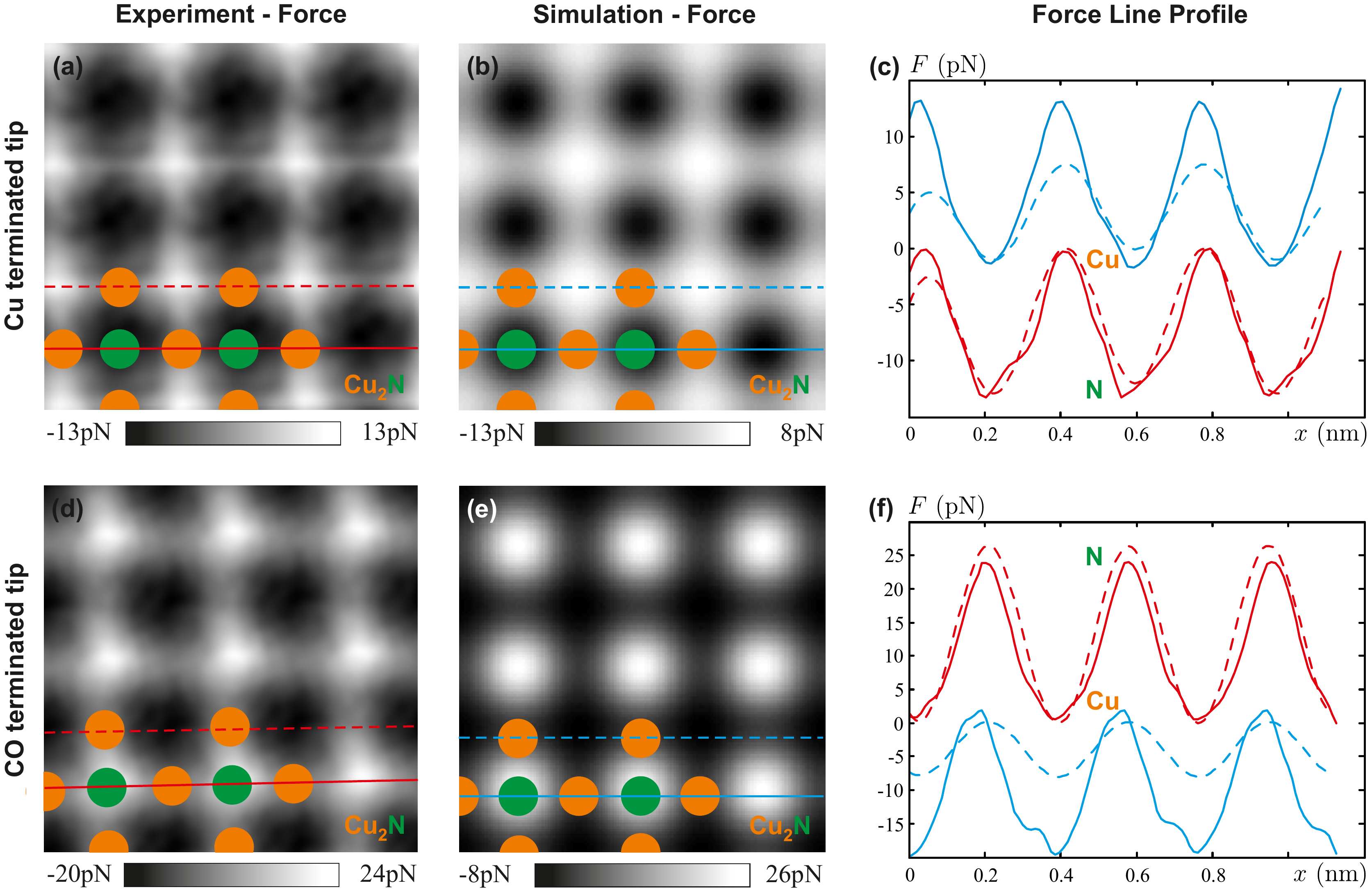}
\caption{Left column: Central area of Fig. \ref{Fig2}b and e (unit cell averaged and $3\times 3$ Gauss filtered). Center column: Calculated electrostatic force between the tip dipole and the point charges of a Cu$_2$N structure, including the second layer Cu atoms (see also Fig. \ref{Fig1}). Simulation for a Cu terminated tip with a dipole moment of 0.874\,D (b) and a CO terminated tip with 0.035\,D (e). In the first two columns, the force at the Cu site was set to zero to emphasize the force difference. Right column: Line profiles from (a), (d) in red and (b), (e) in blue, along N-Cu-N (solid) and Cu-hollow site-Cu (dashed) direction respectively.}
\label{Fig3}
\end{figure*}

From experimental \cite{Choi2008} as well as theoretical work \cite{Yoshimoto2002, Hirjibehedin2007, Soon2008}, the structural parameters and net charges of the Cu$_2$N's unit cell atoms are known. The distance between N and Cu is 183\,pm, N atoms are vertically displaced by 21\,pm with respect to the Cu surface layer and the charged subsurface Cu layer has 197\,pm distance to the Cu surface layer, as indicated in Figs. \ref{Fig1}c-e. Charges were calculated by DFT in Ref. \cite{Hirjibehedin2007} and are given by -1.2\,e for N, +0.5\,e for surface Cu atoms and +0.1\,e for the subsurface Cu (Figs. \ref{Fig1}c-e). Using these numbers, we constructed a 3D point-charge representation of the sample with a size of $7 \times 7$ unit cells, including the subsurface Cu atoms, where all charges are located at the atomic core positions. The resulting electric field was then used to calculate the interaction force with the tip \cite{SM}. The Smoluchowski effect \cite{Smoluchowski1941} causes a charge redistribution on corrugated metal surfaces, which leads to the formation of electric dipoles. For metallic tips, an electric dipole moment is formed pointing towards the sample \cite{Teobaldi2011}, as depicted in Fig. \ref{Fig1}a. For a CO molecule adsorbed on a metal tip, the electric dipole moment compared to gas phase changes in magnitude and sign due to charge transfer to the metal. We recently found experimental \cite{Welker2012, Hofmann2014} evidence that the CO's dipole moment is pointing into the metal, as depicted in Fig. \ref{Fig1}b, which is also supported by some \textit{ab initio} molecular orbital calculations \cite{Pavao1991}. The tip was modeled as a simple dipole having two charges $\pm q$ separated by the dipole distance $d$. 

For Cu terminated tip, we got the best agreement with a charge of $\pm$\,0.13\,e and $d=135$\,pm (close to the atomic radius of Cu) resulting in a dipole moment of $0.182\,\text{e\AA}=0.874\,$D. The dipole points towards the sample, as described before, where the positive charge is located at the core of the front most Cu tip atom. For the calculation, the positive tip charge (Cu core) was placed at a distance of 395\,pm over the charges (cores) of the Cu surface layer. This distance was obtained by adding twice the atomic radius of Cu to the experimental $z_{\text{s-s}}=125$\,pm \cite{SM}. Figure \ref{Fig3}b shows the calculated force between this metal tip dipole and all surface atoms. The force above the Cu site was set to zero to emphasize the force contrast. Line profiles along the N-Cu-N and Cu-hollow site-Cu direction from Fig. \ref{Fig3}b are plotted in Fig. \ref{Fig3}c (blue). The relative interaction contrast is given in Table \ref{table} and fits well to the experimental results. To compare the simulation to the experiment, Fig. \ref{Fig3}a shows a zoom-in of the island in Fig. \ref{Fig2}b (unit cell average over the inner part plus $3\times 3$ Gauss filter) with the respective line profiles in Fig. \ref{Fig3}c (red).

We modeled the CO tip as a dipole pointing away from the sample, where the negative charge sits at the O core and positive charge at the C core. With the CO binding length of 115\,pm as dipole distance $d$ and a charge of $\pm$\,0.03\,e the dipole moment is $0.035\,\text{e\AA}=0.166$\,D. The distance of the O core to the charges of the Cu surface layer was 275\,pm ($z_{\text{s-s}}=80$\,pm plus atomic radii of O, 60\,pm, and Cu, 135\,pm) \cite{SM}. In Fig. \ref{Fig3}e, the calculated force is shown, with the force at the Cu site set to zero. Corresponding line profiles along the two prominent direction are given in Fig. \ref{Fig3}f (blue lines). Also, this simulation gives quantitative agreement to our experimental force contrast (Table \ref{table}). For comparison Fig. \ref{Fig3}d shows a zoom-in of Fig. \ref{Fig2}e (unit cell average over the inner part plus $3\times 3$ Gauss filter), with the corresponding line profiles shown in Fig. \ref{Fig3}f.

This straight-forward electrostatic model reproduces the relative force contrast of both tip terminations very well. The experimentally determined dipole moment of the Cu terminated tip of 0.874\,D is close to the range of previous work for Cr and W tips \cite{Teobaldi2011, Trevethan2012}. Simulation and experiment for the CO terminated tip agree with the theory that the dipole moment points away from the sample. In general the CO's dipole magnitude depends on the chemical nature and geometric structure of the adsorbent \cite{Pavao1991}. Hence, it is not surprising that the fitted value of 0.166\,D varies from CO molecules adsorbed on surfaces \cite{Feng2011, Welker2012, Hofmann2014}. 

The simulation does not account for (attractive) covalent bonds (Cu tips) and Pauli repulsion (CO tips). Both interactions are probably included in the measured overall force and would influence the force contrast and therefore the modeled dipole moment. However, the experimental core-core distances are just a lower boundary (Cu tip: 395\,pm and CO tip: 275\,pm) and are so large that we don't expect a strong contribution neither from covalent bonds nor from Pauli repulsion.

In conclusion, we have shown that atomic resolution on relatively inert surfaces such as Cu$_2$N can also be obtained  by electrostatic multipole forces, not just by covalent bonding forces or Pauli repulsion forces. Richard Feynman already pointed out in his 1939 paper \cite{Feynman1939} that intermolecular forces ultimately have an electrostatic origin. The difference we observed here to the covalent case in atomic imaging of semiconductors is that covalent bonds usually result in massive rearrangements of electronic charge density and subsequently forces in the nanonewton regime, while the forces here are only tens of piconewtons with much less redistribution of electronic charge.

\begin{acknowledgments}
The authors thank the Deutsche Forschungsgemeinschaft for funding within the SFB 689.
\end{acknowledgments}

\clearpage

 \multido{\i=1+1}{4}{
       \begin{figure}
       \center
       \vspace{-2cm}\includegraphics[page=\i,scale=0.9]{Supplemental_Information.pdf}
       \end{figure}
        \newpage
        \pagenumbering{gobble}
    }

\end{document}